\documentclass[preprint,superscriptaddress,aps,prb]{revtex4-2}
\usepackage{amsmath,mathtools}
\usepackage{amssymb}
\usepackage{gensymb}
\usepackage{amsthm}
\usepackage{graphicx}
\usepackage{physics}
\newcommand{\e}{\varepsilon}
\newcommand{\bea}{\begin{eqnarray}}
\newcommand{\eea}{\end{eqnarray}}
\newcommand{\beq}{\begin{equation}}
\newcommand{\eeq}{\end{equation}}

\begin{document}
	
\title{Nematic Phase Transitions in Multilayer Graphene Systems}

\author{R. David Mayrhofer}
\affiliation{School of Physics and Astronomy and William I. Fine Theoretical Physics Institute, University of
Minnesota, Minneapolis, MN 55455, USA}

\author{Andrey V. Chubukov}
\affiliation{School of Physics and Astronomy and William I. Fine Theoretical Physics Institute, University of
Minnesota, Minneapolis, MN 55455, USA}

\date{\today}

\begin{abstract}
Recent experiments on graphene multilayers under displacement field have demonstrated a wide variety of electronically ordered phases, including valley and/or spin polarized phases as well as potentially unconventional superconducting phases. In addition, quantum oscillation measurements in Bernal bilayer graphene and rhombohedral trilayer graphene showed the presence of electronic nematic states. Here, we investigate the emergence of nematic order in these systems, with emphasis on Bernal bilayer graphene, within a self-consistent Hartree--Fock framework, using a realistic band structure with trigonal warping and a dual-gate screened Coulomb interaction. We compute the phase diagram as carrier density and displacement field are varied, and find a sequence of isospin (spin and valley) polarized states consistent with experiment, including partially polarized phases with a large Fermi pockets for isospin-majority carriers and smaller Fermi pockets for isospin-minority carriers. Within these partially isospin polarized states, we identify regions with three $C_3$-symmetric pockets for the minority carriers and regions with only one or two such pockets, implying that the system develops a spontaneous nematic order that breaks $C_3$ symmetry. We find numerically that nematicity emerges near the boundary between fully and partially polarized phases and is controlled by the strength of the screened interaction. We analytically derive a criterion for the nematic order, which agrees with our numerical results.
\end{abstract}
\maketitle

\section{Introduction}
Recent experiments on AB- and ABC-stacked graphene multilayers have revealed a rich landscape of strongly correlated electronic phases. Symmetry-broken metallic and superconducting phases have been observed in systems with two through six layers ~\cite{zhou2021rtg,zhou2021sc,zhou2022bbg,seiler2022,*Seiler2024,barrera2022,han2023,trevol2024,han2024,holleis2025,Auerbach2025,nguyen2025,han2025superconductivity,deng2026}. Bernal bilayer graphene (BBG) and rhombohedral trilayer graphene (RTG) were among the first of these systems to be studied and have emerged as a key platform for correlated electron physics. It provides a setting where strong correlation effects can be systematically explored by varying electron density and the magnitude of the applied electric field, commonly called a displacement  field.  These systems also exhibit superconductivity that appears in a magnetic field  ~\cite{zhou2021rtg,zhou2021sc,zhou2022bbg} or when the system is in proximity to WSe$_2$~\cite{zhang2023,li2024,patterson2025,holleis2025}.

One of the key challenges is to identify and characterize the normal-state phases out of which superconductivity develops. BBG and RTG  offer a controlled setting in which this problem can be addressed. Its low-energy electronic structure is effectively described by a tight-binding model with band minima located around the $K/K'$ points of the Brillouin zone. At low carrier densities, trigonal warping splits the Fermi surface in each valley into three smaller pockets related by $C_3$ symmetry. As density increases, these pockets grow in size until they join together to form a single large Fermi surface (see Fig. \ref{fermi_surface_evolution} below). 

In the absence of interactions, such a system exhibits a four-fold degeneracy associated with spin and valley degrees of freedom~\cite{mccann2006,mccann2006asymmetry,McCann2013,jeil2014}. Electron-electron interactions qualitatively modify this picture. 
A body of experimental work indicates \cite{zhou2021rtg,zhou2022bbg,seiler2022,*Seiler2024,barrera2022,trevol2024,holleis2025}
that the spin-valley degeneracy is lifted in the correlated metallic phases, leading to a sequence of isospin-polarized states. Depending on filling and the magnitude of the displacement field, the system can realize polarized phases in which only one or two spin-valley flavors are occupied, corresponding to quarter- and half-metals, as well as intermediate partially isospin polarized (PIP) states.  These PIP states can be thought of as individually shifting the chemical potentials of the different isospins, creating majority and minority carriers. The chemical potential of the majority carriers is larger, while the chemical potential of the minority carriers is smaller. Because of this, one expects the majority carriers to have a single larger Fermi surface per isospin and the minority carriers to have three smaller Fermi surfaces, again per isospin. 
   [Two larger and $2 \times 3$ smaller Fermi surfaces in the PIP phase next to a half metal and one larger and three smaller Fermi surfaces in the PIP phase next to a quarter metal]. 
Theoretical studies have largely identified the origin of these isospin-symmetry-broken phases, emphasizing the role of exchange interactions and the momentum dependence of the Coulomb interaction in stabilizing spin- and valley-orders
~\cite{chichinadze2022,*chichinadze2022letters,haoyu2023,xie2023,lee2024,koh2024rtg,koh2024bbg,wang2024electrical,friedlan2025,mayrhofer2025,martinez2025}. 
A number of works have analyzed possible superconducting instabilities in proximity to these ordered states~\cite{ghazaryan2021,ghazaryan2023,pantaleon2023,dong2023transformer,dong2023signatures,dong2023,raines2026twovalley,raines2026}.

\begin{figure}[h]
	\begin{center}
		\includegraphics[scale=.5]{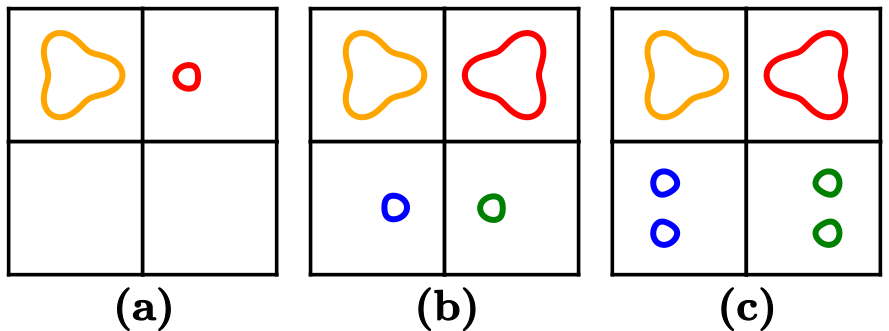}
		\caption{Schematic plots of the experimentally observed phases where the minority carriers break the $C_3$ symmetry by selection of one or two of the pockets. Both (a) and (b) were observed in Refs. ~\cite{zhou2022bbg,zhang2023} and (c) was observed in Ref. \cite{holleis2025}}
		\label{experimental_phases}
	\end{center}
\end{figure}

The goal of this communication is to provide theoretical understanding of other, more subtle aspects of quantum oscillation measurements. Namely the measurements reported in Refs.~\cite{zhou2022bbg,zhang2023,holleis2025} indicate that in portions of PIP states the minority carriers break the $C_3$ symmetry and reside in only one or two pockets. Specifically, Refs.~\cite{zhou2022bbg,zhang2023} found that 
 the minority carriers form a single small Fermi pocket per isospin in the PIP states next to both a quarter-metal   and half-metal state, and Ref. \cite{holleis2025} found two Fermi pockets per isospin for 
 the minority carriers in the PIP state next to a half metal.  We depict these states  in Fig. \ref{experimental_phases}. 
This reduction in the number of occupied pockets implies a breaking of the $C_3$ rotational symmetry of the Fermi surface, and is interpreted as evidence for electronic nematic phases.  To the best of our knowledge, there is no experimental evidence of partially polarized nematic states, e.g., the state with two larger and one smaller pocket for minority carriers. In RTG, there is also evidence for  nematic states~\cite{zhou2021rtg,trevol2024,han2024topology}, although it is not entirely clear whether there is a reduction in the number of pockets for the minority carriers in the same way as in BBG.

In this work, we investigate a spontaneous $C_3-$ breaking by performing a comprehensive theoretical study of interacting electrons in BBG, with a particular focus on the emergence of nematic states.  We consider the tight-binding model that includes trigonal warping. We employ a self-consistent Hartree–Fock approach and consider 
  the $q=0$ orders, i.e. the valley and/or spin polarized states. The interaction is modeled as a dual-gate screened Coulomb potential, allowing us to incorporate the effects of metallic gates and dielectric environment in a realistic manner. By systematically varying the carrier density, displacement field (an electric field 
 that creates an electrostatic potential difference between the two layers), gate distance, and dielectric constant, we construct detailed phase diagrams that can be compared to experiment. 

\begin{figure}[h]
	\begin{center}
		\includegraphics[scale=.65]{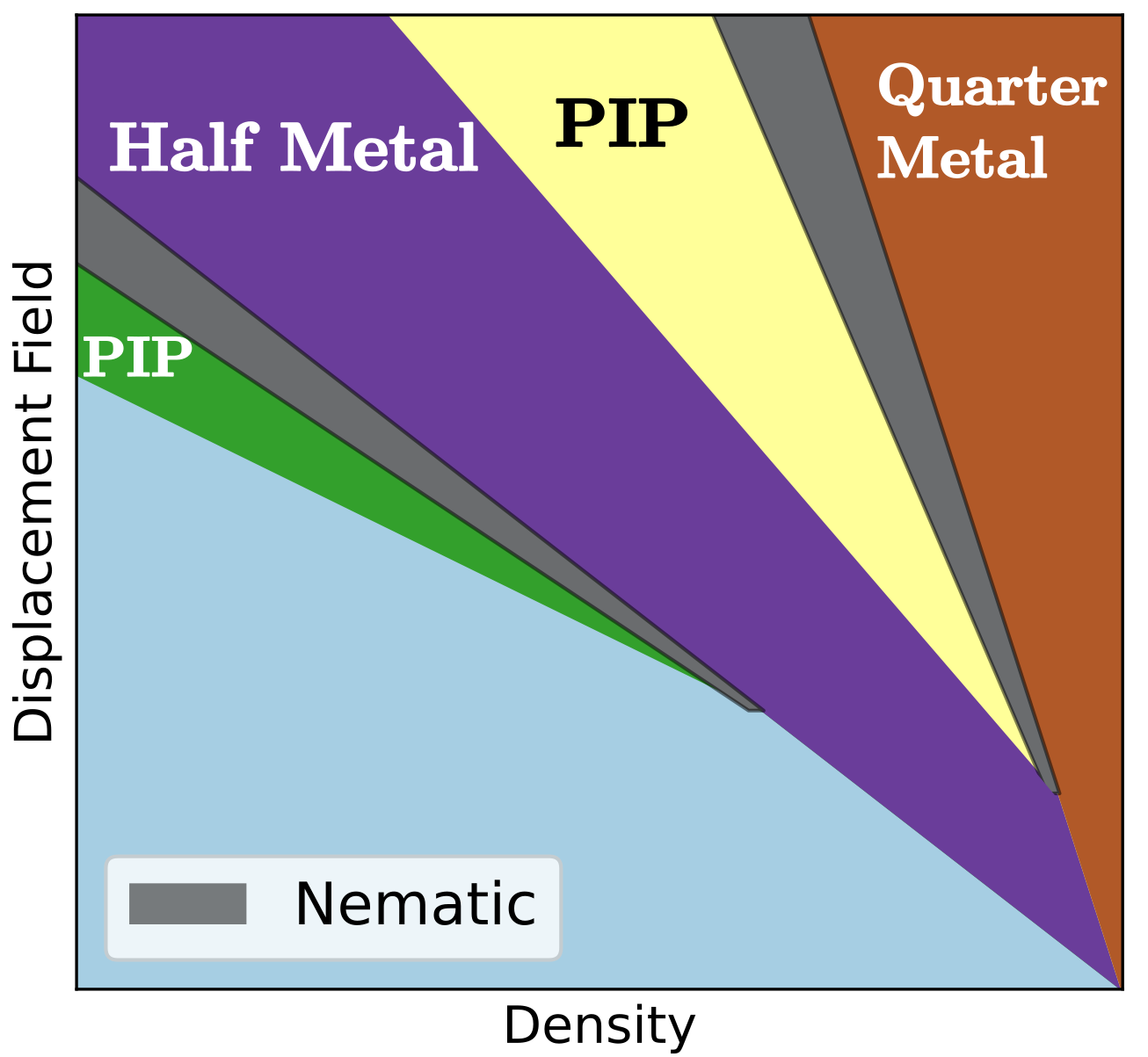}
		\caption{Schematic theoretical phase diagram.  The nematic phases (gray regions) are located in the PIP phases next to their 
 boundaries to either a half-metal or a quarter-metal.   We specify the two PIP phases later in the text.  
 We do not depict the three-quarters metal here, as there are no nematic states that appear in the vicinity of this state.}
		\label{schematic_phase_diagram}
	\end{center}
\end{figure}

Our results show that nematic states do develop within this model. In particular, we find that when the system is partially isospin polarized, the Fermi surfaces of the minority isospin can spontaneously transition from a three-pocket configuration to a one or two pocket configuration. This transition occurs near the boundaries between fully polarized and partially polarized phases, when the occupation of the minority carriers is rather small.  We argue below that the smallness of the pockets increases the effects of the interaction.   
We show the schematic phase diagram with nematic phases in Fig. \ref{schematic_phase_diagram}.

To gain further insight into the origin of this behavior, we complement our numerical analysis with an analytical treatment of the low-density regime. Focusing on the three-pocket structure within a single valley, we construct a simplified model in which the pockets are assumed to have a parabolic dispersion, and the intra-pocket interaction is approximated as momentum-independent. Within this framework, we derive analytically a criterion for the onset of nematic order, which captures the competition between interaction energy and the kinetic energy cost associated with redistributing carriers among pockets. This analytical result provides an interpretation of our numerical findings and clarifies the conditions under which nematicity is favored. 

 We view our study as an extension of previous works on spin/valley orders in multi-valley systems.  Some previous works on spin/valley systems ~\cite{raines2024stoner,raines2024isospin,*raines2024unconventional,scholle2026} considered models with momentum-independent Hubbard-type interactions. These models display valley and spin polarization but no nematic order.  
Other works have incorporated momentum-dependent interactions and investigated the general emergence of nematicity in these systems~\cite{chunli2023,dong2023momentum,pantaleon2023,koh2024rtg,koh2024bbg,martinez2025}.
These studies have primarily focused on identifying and characterizing nematic phases rather than determining the conditions under which they form. In addition, Ref.~\cite{calvera2024nematicity} studied a nematic transition in the context of a valley polarized state, where the nematicity arises due to the differences in the Fermi surfaces at the $K$ and $K'$ points in the Brillouin zone. 
In contrast, we systematically analyze the conditions for nematicity within a microscopic model and the Fermi surface structure of the resulting nematic states, with particular emphasis on phases in which the number of occupied pockets for minority carriers is reduced below a nematic transition.
We further derive an analytical criterion for the onset of nematic order in terms of the interaction strength.

The rest of the paper is structured as follows. In Sec.~\ref{background}, we introduce the single-particle Hamiltonian for BBG and specify the form of the screened Coulomb interaction. We also outline the Hartree-Fock procedure used to compute the ground-state energy and describe the analytical model for the nematic instability in the low-density limit. In Sec.~\ref{results}, we present the resulting phase diagrams as functions of dielectric constant and gate distance, and analyze the evolution of spin-valley polarization and nematic order. We show that the analytical criterion for nematicity is in good agreement with the numerical results. Finally, in Section \ref{conclusions}, we present our conclusions.

\section{Background and Formalism}
\label{background}
\subsection{BBG Hamiltonian}
We begin by writing the tight-binding model of BBG,
\begin{align}
\label{single_particle} H_{s} = \sum_{\vb k,\sigma,\tau,l,l'}\psi^{\dagger}_{\vb k,\sigma,\tau,l'}H(\vb k)_{ll'}\psi_{\vb k,\sigma,\tau,l},
\end{align}
where $\sigma$ and $\tau$ are the spin and valley indices while $l,l'$ and are sublattice/layer indices, with $l_1$ $(l_2)$ for the $A$ $(B)$ sublattice of layer 1, and $l_3$ $(l_4)$ for the $A$ $(B)$ sublattice of layer 2. The matrix H is defined as \cite{McCann2013,jeil2014}
\begin{align}
\label{matrix}
H(\vb k)_{l l'} = \begin{pmatrix}
D/2          & \gamma_0 f(\vb k)  & -\gamma_4 f(\vb k)              & -\gamma_3 f^*(\vb k) \\
\gamma_0 f^*(\vb k) &  \delta+D/2           & \gamma_1    & -\gamma_4  f(\vb k)         \\
-\gamma_4 f(\vb k)^*              & \gamma_1       & \delta-D/2       & \gamma_0 f(\vb k)   \\
-\gamma_3 f(\vb k)   & -\gamma_4 f(\vb k)^*    & \gamma_0 f^*(\vb k) & -D/2             
\end{pmatrix}_{l l'},
\end{align}
where $D$ is the displacement field, which is diagonal on the sublattice index but changes sign between different layers. The $\gamma_i$ are the hopping parameters, $\delta$ is the AB-sublattice potential difference, and $f(\vb k)$ is the hexagonal lattice form factor associated with nearest-neighbor hopping. We wrote this as
\begin{align}
f(\vb k) = -e^{i k_y a/\sqrt{3}} - 2 e^{-i k_y a/2\sqrt{3}} \cos \frac{k_x a}{2}.
\end{align}
We note that this single-particle Hamiltonian may be modified by spin-orbit coupling. In BBG, the spin-orbit coupling is small (around 10-100$\mu$eV), and can be neglected~\footnote{The spin orbit coupling may be enhanced by placing the sample in proximity of a TMD like WSe$_2$. In this work, we will consider the case where the sample is not in the proximity of any TMD, and the spin-orbit coupling can be safely neglected}. Out of these tight-binding parameters, $\gamma_0$ and $\gamma_1$, play the most prominent role~\cite{McCann2013}. In particular, $\gamma_3$ gives the leading order trigonal warping term, and must be included to accurately reproduce several features observed in the experimental phase diagram~\cite{mayrhofer2025}. The terms $\gamma_4$ and $\delta$ correspond to subleading corrections to the dispersion. 

We determine the dispersion close to the $K/K'$ points. We calculate the dispersion to order $k^4$ in Appendix \ref{effective_mass} by expanding in $D/\gamma_1$, $\gamma_3/\gamma_0$, $\gamma_4/\gamma_0$, and $\delta/\gamma_1$ and only keeping leading order terms. We have
\begin{align}
\label{disp} \e_{\vb k,\tau} = \gamma_1  \left(-\frac{D}{2 \gamma_1} +  \left(\frac{D}{\gamma_1} - \frac{\gamma_3^2\gamma_1}{\gamma_0^2D} + 2\frac{\gamma_4}{\gamma_0} + \frac{\delta}{\gamma_1} \right)z^2 - 2\frac{\gamma_3\gamma_1}{\gamma_0 D} \tau \cos(3\theta)z^3-\left( \frac{\gamma_1}{D} +\frac{\gamma_3}{\gamma_0} + 4\frac{\gamma_4}{\gamma_0} + 3 \frac{\delta}{\gamma_1} \right)z^4 \right),
\end{align}
where $\tau=\pm 1$ is the valley index, $z = \frac{\sqrt{3}\gamma_0}{2 \gamma_1} a k$, and $k$ is measured from the $K/K'$ points. 

\begin{figure}[h]
	\begin{center}
		\includegraphics[scale=.65]{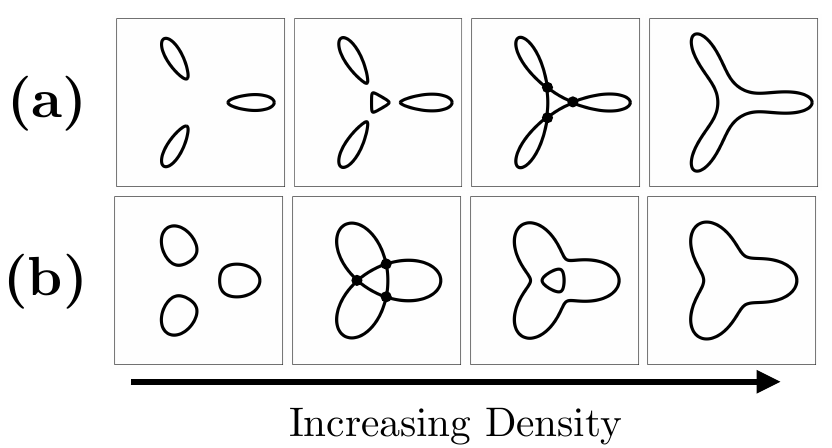}
		\caption{A schematic depiction of the evolution of the Fermi surface of a single isospin in BBG with a displacement field (a) $D<D_c$ and (b) $D>D_c$.}
		\label{fermi_surface_evolution}
	\end{center}
\end{figure}

The evolution of the Fermi surface as density is varied is demonstrated in Fig. \ref{fermi_surface_evolution}. It is obtained from Eq. \ref{disp} by solving for $\e{\vb k} = \mu$, where $\mu$ is the density dependent chemical potential. When the displacement field is smaller than some critical value, $D_c$, the evolution goes as in Fig. \ref{fermi_surface_evolution}a, whereas if $D>D_c$, the Fermi surfaces evolve as in Fig. \ref{fermi_surface_evolution}b. From Eq. (\ref{disp}), we find 
\begin{align}
D_c = \frac{\sqrt{\left(\gamma _0 \delta +2 \gamma _1 \gamma _4\right)^2+4 \gamma _1^2 \gamma _3^2}-\gamma _0 \delta -2 \gamma _1 \gamma _4}{2 \gamma _0} \sim 20\text{meV}.
\end{align}

We then consider the interacting part of the Hamiltonian. Here, we take the electron-electron interactions to be modeled by the dual-gate screened Coulomb interaction,
\begin{align}
V_C(\vb q) = \frac{e^2}{2\e_{r}\e_{0}q} \tanh \left(q d\right),
\end{align}
where $d$ is the distance from the gates to the BBG sample, $\e_r$ is relative permittivity, and $\e_0$ is the permittivity of free space. We write the interacting part of the Hamiltonian as
\begin{align}
H_I = \frac{1}{2A}\sum_{\vb k, \vb p,\vb q}\sum_{\sigma \sigma' \tau \tau' l l'} V_C(\vb q) \psi^\dagger_{\vb k+\vb q,\sigma, \tau,l} \psi^{\dagger}_{\vb p- \vb q,\sigma',\tau',l'}\psi_{\vb p,\sigma',\tau',l'}\psi_{\vb k,\sigma,\tau,l},
\end{align}
where $A$ is the area of the system. When transforming the single particle Hamiltonian from the sublattice/layer basis to the band basis, coherence factors are introduced into the interaction Hamiltonian. Formally, this transforms each of the operators as $\psi_{\vb k,\sigma,\tau,l} = \sum_{i} U^{\tau}_{l i}(\vb k) c_{\vb k,\sigma,\tau,i}$, where each $U^{\tau}(\vb k)$ is a $4\times4$ unitary matrix. 

In addition, each of the bands are well separated from each other when the system is under a displacement field. Since we consider the case of hole doping at small densities, we single out the band that crosses the Fermi level and discard the rest. Then, the total Hamiltonian takes the form
\begin{align}
\label{hamiltonian} H &= \sum_{\vb k,\tau} \e_{\vb k,\tau} c^{\dagger}_{\vb k,\sigma,\tau}c_{\vb k,\sigma,\tau} + \frac{1}{2A}\sum_{\vb k, \vb p,\vb q}\sum_{\sigma,\sigma',\tau,\tau'} V_{eff,\tau,\tau'}(\vb k,\vb p,\vb q) c^\dagger_{\vb k+ \vb q,\sigma,\tau} c^{\dagger}_{\vb p- \vb q,\sigma',\tau'}c_{\vb p,\sigma',\tau'}c_{\vb k,\sigma,\tau},
\end{align}
where we have written
\begin{align}
V_{eff,\tau,\tau'}(\vb k,\vb p,\vb q) = \sum_{l,l'} V_C(\vb q) a^*_{\tau,l}(\vb k+ \vb q)a^*_{\tau',l'}(\vb p- \vb q)a_{\tau',l'}(\vb p)a_{\tau,l}(\vb k).
\label{ss:1}
\end{align}
We define the coherence factors as $a_{\tau,l}(\vb k) = U^{\tau}_{l2}(\vb k)$ and have suppressed the band index on the creation and annihilation operators. 

\subsection{Spin and valley orders: Hartree-Fock analysis}
We first analyze spin and valley orders, which we treat as the primary orders in our system.  These orders give rise to half-metal and quarter-metal states with two or one pockets for majority carriers, respectively and no Fermi surfaces for fermions from other valley/spin projection, and to four types of PIP states (see schematic Fig. \ref{fermi_surface_evolution} and more detailed Fig. \ref{fermi_surfaces}). These PIP states contain three small pockets for minority carriers for a given valley and spin projection.
We then analyze subleading instabilities within these PIP states towards $C_3$ breaking  nematic orders with one or two pockets instead of three. 

We restrict our attention to the $q=0$ orders, that is spin and valley polarized states, with order parameters
 of the form $c^\dagger_{\vb k,\sigma,\tau} c_{\vb k,\sigma,\tau}$.    An extension to charge and spin inter-valley coherent orders (the ones with momentum $K-K'$) is straightforward, but requires a separate analysis.   

In the Hartree-Fock calculation of valley and spin orders,  one introduces fermionic bilinears  corresponding to valley polarization and ferromagnetism   
\begin{align}
\Delta_{VP}(\vb k) &= \sum_{\sigma,\tau} \Delta_{\sigma,\tau}(\vb k)\text{sgn}(\tau) \\ 
\Delta_{FM}(\vb k) &= \sum_{\sigma,\tau} \Delta_{\sigma,\tau}(\vb k)\text{sgn}(\sigma)
\end{align}
where 
\begin{align}
\label{order_parameter} \Delta_{\sigma,\tau}(\vb k) = \frac{1}{A}\sum_{\vb p} V_{eff,\tau,\tau}(\vb k, \vb p,\vb p - \vb k)  \langle c^\dagger_{\vb p,\sigma,\tau}c_{\vb p,\sigma,\tau} \rangle.
\end{align}
and use then to decouple  the 4-fermion interaction term into the  $O(\Delta^2)$ term and the  $O(\Delta)$  term  bilinear in fermions, which one adds to the kinetic energy.
We find that  the Hartree term contributes only a constant shift to the energy, which can be absorbed into the chemical potential. The Fock term is
\begin{eqnarray}
H_I^{Fock}&=& \frac{1}{2A}\sum_{k,p}\sum_{\sigma,\tau} V_{eff,\tau,\tau}(\vb k, \vb p,\vb p - \vb k) \left(\langle c^\dagger_{\vb k,\sigma,\tau} c_{\vb k,\sigma,\tau} \rangle \langle c^\dagger_{\vb p,\sigma,\tau}c_{\vb p,\sigma,\tau} \rangle - 2 \langle c^\dagger_{\vb p,\sigma,\tau}c_{\vb p,\sigma,\tau} \rangle c^\dagger_{\vb k,\sigma,\tau}c_{\vb k,\sigma,\tau} \right) \nonumber \\
&=&   \sum_{k,\sigma,\tau}  \Delta_{\sigma,\tau}(\vb k )  \left( \frac{1}{2} \langle c^\dagger_{\vb k,\sigma,\tau} c_{\vb k,\sigma,\tau} \rangle - c^\dagger_{\vb k,\sigma,\tau} c_{\vb k,\sigma,\tau} \right)
\end{eqnarray}
Within this approximation, the 
Hamiltonian of the system is
\begin{align}
H= \sum_{k,\sigma,\tau} \left( \frac{1}{2} \Delta_{\sigma,\tau}(\vb k ) \langle c^\dagger_{\vb k,\sigma,\tau} c_{\vb k,\sigma,\tau} \rangle + \left( \e_{\vb k,\sigma,\tau} - \Delta_{\sigma,\tau}(\vb k) \right) c^\dagger_{\vb k,\sigma,\tau} c_{\vb k,\sigma,\tau} \right)
\end{align}
Then, the energy of the system is
\begin{align}
E = \langle H\rangle = \sum_{\vb k,\tau}\left( \e_{\vb k,\tau} -\frac{1}{2}\Delta_{\sigma,\tau}(\vb k) \right) n_F\left(\e_{\vb k,\tau} - \Delta_{\sigma,\tau}(\vb k )-\mu \right),
\end{align}
where $n_F$  is the Fermi function. 
The quantities $\Delta_{\sigma,\tau}(\vb k)$ are obtained by self-consistently solving Eq.~(\ref{order_parameter}), which is equivalent to minimizing the total energy with respect to the order parameters.
    
The resulting solutions determine the occupation of the different isospin flavors. When only a single isospin is occupied, the system realizes a quarter-metal state. Similarly, occupation of two or three isospins corresponds to half-metal and three-quarter-metal states, respectively. States with unequal occupation among multiple isospins are PIP states.  We follow Ref.~\cite{martinez2025}, we denote these states as PIP$_{i,j}$, where $i$ and $j$ are the numbers of majority and minority isospins
(see Fig. \ref{fermi_surfaces}b). 

\subsection{Nematic order: Three Pocket Model}
\label{three_pockets}
Here we present a theory to describe the breaking of the $C_3$ symmetry for minority carriers. To identify the conditions under which this occurs, we focus on a single minority isospin. Since the minority carriers are intrinsically at low density in the PIP phases, their electronic structure consists of three small pockets located either near  $K$ or near $K'$. The dispersion of each of these pockets will be approximately parabolic, forming an elliptical Fermi surface. We then incorporate interactions between carriers within and between these pockets to construct the full interacting model. We assume that coherence factors play a minor role here because all relevant momenta are close and approximate $V_{eff} (k,p,q)$ in Eq. (\ref{ss:1}) by $V_C (q)$
The Hamiltonian for a single isospin is then be written as
\begin{align}
\label{isospin_hamiltonian} H_{minority} = \sum_{\vb k,i} \e_{\vb k,i} d^{\dagger}_{\vb k,i} d_{\vb k,i} +\frac{1}{2A} \sum_{i,j}\sum_{\vb k,\vb p,\vb q} V_C  (\vb q) d^{\dagger}_{\vb k+\vb q,i}d^{\dagger}_{\vb p-\vb q,j}d_{\vb p,j}d_{\vb k,i},
\end{align}
where $d_{\vb k,i}$ is the annihilation operator of particle with momentum $\vb k$ in pocket $i$, and $\e_{\vb k,i} =\frac{(k_x-k_{x,i})^2}{2m_x^*} + \frac{(k_y-k_{y,i})^2}{2m_y^*}$ is the parabolic dispersion and the $\vb k_i$ are the center of each of the three pockets. 

The goal is to determine under what conditions it will be favorable for one or two of the three original pockets to be selected. We determine these conditions again performing a mean field calculation on this Hamiltonian. Here, we will assume that the size of the pockets is small relative to the screening length, so that the intra-pocket interaction in the Fock term can be approximated as $V_C(\vb k-\vb p) \sim V_C(q)$, where $q$ is a characteristic momentum transfer of particles within a pocket.
In addition, we assume that center of the pockets does not move as density increases, so the inter-pocket interaction is $V_C(Q)$, where $Q$ is the distance between the centers of the pockets.
 We then write down order parameters for the system, 
\begin{align}
\Delta_1 &= \frac{\tilde V}{3A} \sum_{\vb k} \left( 2 \langle d^{\dagger}_{\vb k,1} d_{\vb k,1} \rangle - \langle d^{\dagger}_{\vb k,2} d_{\vb k,2} \rangle - \langle d^{\dagger}_{\vb k,3} d_{\vb k,3} \rangle\right)\\
\Delta_2 &= \frac{\tilde V}{3A} \sum_{\vb k} \left( 2 \langle d^{\dagger}_{\vb k,2} d_{\vb k,2} \rangle - \langle d^{\dagger}_{\vb k,1} d_{\vb k,1} \rangle - \langle d^{\dagger}_{\vb k,3} d_{\vb k,3} \rangle\right)\\
\Delta_3 &= \frac{\tilde V}{3A} \sum_{\vb k} \left( 2 \langle d^{\dagger}_{\vb k,3} d_{\vb k,3} \rangle - \langle d^{\dagger}_{\vb k,1} d_{\vb k,1} \rangle - \langle d^{\dagger}_{\vb k,2} d_{\vb k,2} \rangle\right),
\end{align}
where $\tilde V = V_C(q)  - V_C (Q)$. We have $\Delta_1 + \Delta_2 + \Delta_3 = 0$ as is required by particle conservation, so we simplify the above relations by writing 
\begin{align}
\Delta_3 = - \Delta_1 - \Delta_2.
\end{align}
With these definitions of the order parameters, we then perform mean field calculations on Eq. (\ref{isospin_hamiltonian}). Doing so, we find that the energy per isospin is
\begin{align}
\frac{E}{A} = \frac{1 - \nu_F \tilde V}{\nu_F\tilde V^2} \left(\Delta_1^2 + \Delta_2^2 + \Delta_1 \Delta_2\right) + \cdots
\label{tt_1}
\end{align}
where $\nu_F = \sqrt{m_x^* m_y^*}/2\pi$ is the density of states at the Fermi surface and the ellipses stand for terms that do not depend on $\Delta_{1,2}$ and therefore do not influence the onset of nematic order. We also note that this is the full expression for the energy, not the expansion in $\Delta_i$.  Since the dispersion is parabolic and the system is 2D, there are no quartic and higher-order terms in $\Delta_i$ in the energy\cite{raines2024unconventional,raines2024isospin}. The condition for the appearance of finite $\Delta_i$ is obviously
\begin{align}
\nu_F \tilde V >1.
\end{align}
We solve the self consistent equations on $\Delta_1$ and $\Delta_2$ for $ \nu_F \tilde V >1$. To do so, we note that $\langle d^{\dagger}_{\vb k,i} d_{\vb k,i} \rangle = n_F(\e_{\vb k,i}-\Delta_i-\mu)$. 
The two coupled equations for $\Delta_1$ and $\Delta_2$ are then  
\begin{align}
\Delta_1 &= \frac{\tilde V}{3A} \sum_{\vb k} \left( 2 n_F(\e_{\vb k,1}-\Delta_1-\mu) - n_F(\e_{\vb k,2}-\Delta_2-\mu) - n_F(\e_{\vb k,3}+\Delta_1+\Delta_2-\mu)\right)\\
\Delta_2 &= \frac{\tilde V}{3A} \sum_{\vb k} \left( 2 n_F(\e_{\vb k,1}-\Delta_2-\mu) - n_F(\e_{\vb k,2}-\Delta_1-\mu) - n_F(\e_{\vb k,3}+\Delta_1+\Delta_2-\mu)\right).
\end{align}
In addition to these two equations, there is another one for total particle number in the isospin $N = nA $  that  must be conserved.
This equation is 
\begin{align}
n = \frac{1}{A} \sum_{\vb k,i} n_F(\e_{\vb k,i}-\Delta_i-\mu).
\end{align}
For $\Delta_i =0$,  $(1/A) \sum_{\vb k} n_F(\e_{\vb k,i}-\mu) = \nu_F \mu$,  for  a finite  $\Delta_i$, $(1/A) \sum_{\vb k} n_F(\e_{\vb k,i}-\Delta_i -\mu) = \nu_F (\mu + \Delta_i)$ if $\mu + \Delta_i >0$ and zero if $\mu + \Delta_i <0$.
Solving the nonlinear self consistent equations while conserving particle number, we find three solutions for the order parameters. The first is the trivial $C_3$-preserving 
solution $\Delta_1 = \Delta_2 = \Delta_3 = 0$.
The second is the two pocket configuration, where one pocket is empty and two are occupied. 
Without loss of generality, assume and then verify that the third pocket is empty, i.e.,  that $ n_F(\e_{\vb k,3}+\Delta_1+\Delta_2-\mu)=0$.  
The system of equations is then
\begin{align}
\Delta_1 &= \frac{\nu_F \tilde V}{3} \left( 2 \Delta_1 - \Delta_2 + \mu \right) \\
\Delta_2 &= \frac{\nu_F \tilde V}{3} \left( 2 \Delta_2 - \Delta_1 + \mu \right) \\ 
n &= \nu_F  \left(\Delta_1 + \Delta_2 + 2 \mu \right)
\end{align}
Solving this system of equations, we find $\Delta_1 = \Delta_2 = \frac{\tilde V n }{6}$ and $\mu = \frac{n}{2 \nu_F} - \frac{\tilde V n }{6}$.
For these values, the effective chemical potentials for pockets $1$ and $2$ are   $\Delta_1 + \mu = \Delta_2 + \mu =  \frac{n}{2 \nu_F} >0$, i.e., these two pockets are occupied, while for the third pocket, the effective chemical potential   is $-(\Delta_1+ \Delta_2) + \mu = \frac{n}{2 \nu_F} (1 -\nu_F {\tilde V})$.  It  is negative for $\nu_F {\tilde V} >1$, when energy consideration favors a nematic order.  This confirms our assumption that the third pocket is fully depleted. 
 
The last solution is the one pocket configuration, where two pockets are empty and one pocket is occupied. Similarly to the two pocket configuration, we assume, without loss of generality, that second and third pockets are fully depleted, i.e., 
 $n_F(\e_{\vb k,2}-\Delta_2-\mu) = n_F(\e_{\vb k,3}+\Delta_1+\Delta_2-\mu) =0$. The system of equations  then  reduces to   
\begin{align}
\Delta_1 &= \frac{2 \nu_F \tilde V}{3} \left(\Delta_1 + \mu \right) \\
\Delta_2 &= -\frac{\nu_F \tilde V}{3} \left(\Delta_1 + \mu \right) \\ 
n &= \nu_F \left(\Delta_1 + \mu \right).
\end{align}
The solution to this system is $ \Delta_1 = \frac{2 \tilde Vn}{3}$, $\Delta_2 = -\frac{\tilde Vn}{3}$ and $\mu = \frac{n}{\nu_F}-\frac{2 \tilde Vn}{3}$. We see that $\Delta_3 = -(\Delta_1 +\Delta_2) = -\frac{\tilde Vn}{3} = \Delta_2$.  The effective chemical potential for pocket $1$ is then positive, $\Delta_1 + \mu = \frac{n}{\nu_F}>0$, i.e., this pocket is occupied, while the effective chemical potentials for pockets $2$ and $3$,  $\Delta_2 + \mu$ and $-(\Delta_1+ \Delta_2) + \mu$ are both equal to  $\frac{n}{\nu_F} \left(1 - \nu_F \tilde V  \right)$ and are negative for $\nu_F \tilde V >1$.  Hence these pockets are fully depleted, as we anticipated.   
 We emphasize that (i) there are no solutions with partial occupation of the three pockets (e.g., two larger and one smaller or one larger and two smaller) and (ii) the nematic susceptibility diverges on approaching $\nu_F \tilde V  =1$ from below, as evidenced from Eq. (\ref{tt_1}).   

Using the solutions for the one pocket configuration and the two pocket configuration, we determine the energy difference between the ordered states and the trivial state, which are
\begin{align}
\frac{\delta E_{1}}{A} =  \frac{n^2\left(1 - \nu_F \tilde V\right)}{3\nu_F}\\
\frac{\delta E_{2}}{A} =  \frac{n^2\left(1 - \nu_F \tilde V\right)}{12\nu_F},
\end{align}
where the $\delta E_i$ correspond the energy difference between the normal state and the state with $i$ pockets. 
We see that $\delta E_{1,2}$ is negative for  $\nu_F \tilde V>1$. We also  see that  $\delta E_{1}< \delta E_{2}$, so the one pocket configuration is favored over the two pocket one. On a more careful look, we find that this  holds only for a parabolic dispersion.  Using the full dispersion for BBG, we find numerically (see below) that the initial transition from a $C_3$-symmetric state is into two pocket configuration, and one-pocket configuration emerges at larger $\nu_F \tilde V$.

To compare with the numerical results, we recall that 
\begin{align}
\label{nematicity_condition} \nu_F \tilde V = \nu_F\left(V_C(q) - V_C(Q)\right) = \frac{\nu_F e^2d}{2 \e_r \e_0}\left(
\frac{\tanh(qd)}{qd} - \frac{\tanh(dQ)}{dQ}\right)
\end{align}
 where  $q$ as the characteristic momentum transfer of particles within a pocket, $Q$ is the distance between the centra of the pockets, and $d$ is the distance from the gates to BBG.   
To determine whether the condition $\nu_F \tilde V >1$ holds, we need to express  both $\nu_F$ and $Q$ 
 in terms of experimentally determined parameters of the tight-binding model. 
For an elliptical Fermi surface in 2D, the density of states is $\nu_F = \sqrt{m_x^{*} m_y^*}/2\pi$. We calculate the effective mass alongside $Q$, in Appendix \ref{effective_mass}. 
We find that for small $q$, i.e., when  the pockets for minority carriers are small within PIP phase and  $\tanh(qd)/qd \approx 1$, the condition 
 $ \nu_F \tilde V >1$ holds.   For larger $q$, $\tanh(qd)/qd$ decreases and $\nu_F \tilde V$ becomes smaller than one.  
The outcome here is that the best condition for nematicity is when minority isospins  are at the lowest densities. This occurs when the PIP phases are close to fully polarized states. We will show in the next section that in the numerical analysis we indeed find nematic phases in PIP states close to the boundaries of the fully polarized states.   

We further note that this behavior is specific to two-dimensional systems. In three dimensions, the density of states decreases as the carrier density is reduced, whereas in two dimensions it approaches a constant at low densities. As a result, in 2D the criterion for nematicity at low densities is controlled primarily by the interaction at small momentum transfer.
\section{Results}
\label{results}
Here, we present the computed phase diagrams for BBG within Hartree-Fock as number density and displacement field strength are varied. To construct these phase diagrams, we solve Eq. \ref{order_parameter} for each $\Delta_{\sigma,\tau}(\vb k)$. The self-consistent procedure is initialized with multiple distinct starting configurations, each chosen to approximate a different candidate phase (e.g., quarter metal, half metal, etc.). The solution is obtained iteratively while conserving total number of particles at each step. Among the converged solutions, we keep the state that minimizes the free energy. We use the full band structure, obtained by numerically diagonalizing Eq. (\ref{matrix}). In experiments, the system is lightly hole doped, so that the only relevant band is the higher energy hole band, as the remaining bands are separated by sizable gaps..
In the numerics, we discard all bands except this one. This procedure is carried out for each point in the phase diagram to generate the phase diagrams presented below.

In this system, it is sufficient to restrict momenta to be close to the $K/K'$ points. We implement this by introducing a momentum cutoff $\Lambda$ around the $K/K'$ points, and only keep momenta smaller than that cutoff. For each displacement field and density, we choose the momentum cutoff so that it includes the entire Fermi surface without extending to unnecessarily large momenta. To do this, we estimate the largest possible Fermi momentum $k_{F,\max}$, which occurs when all particles occupy a single isospin, and then set the cutoff to be larger, $\Lambda = 1.2 k_{F,\max}$. In addition, the momentum grid is also constructed adaptively, such that the total number of momentum points remain approximately $2\times 10^4$. This relatively high resolution is important for accurately resolving ordered phases, particularly at low densities. Throughout, we work at a finite temperature $T = 3\times 10^{-2}$ meV ($\sim 30$ mK), comparable to experimental conditions. 
For all numerics, we will be using the values for the $\gamma_i$ and $\delta$ obtained from \textit{ab initio} calculations in Ref. \cite{jeil2014}:
\begin{align}
\label{parameters} \gamma_0 = 2610 \text{meV}, \, \gamma_1 = 361 \text{meV}, \, \gamma_3 = 283 \text{meV}, \, \gamma_4 = 138 \text{meV}, \delta = 15 \text{meV}.
\end{align}
These parameters match reasonably well against experimentally determined values for the tight-binding parameters, see Ref. \cite{jeil2014}. 

\begin{figure}[h]
	\begin{center}
		\includegraphics[scale=.25]{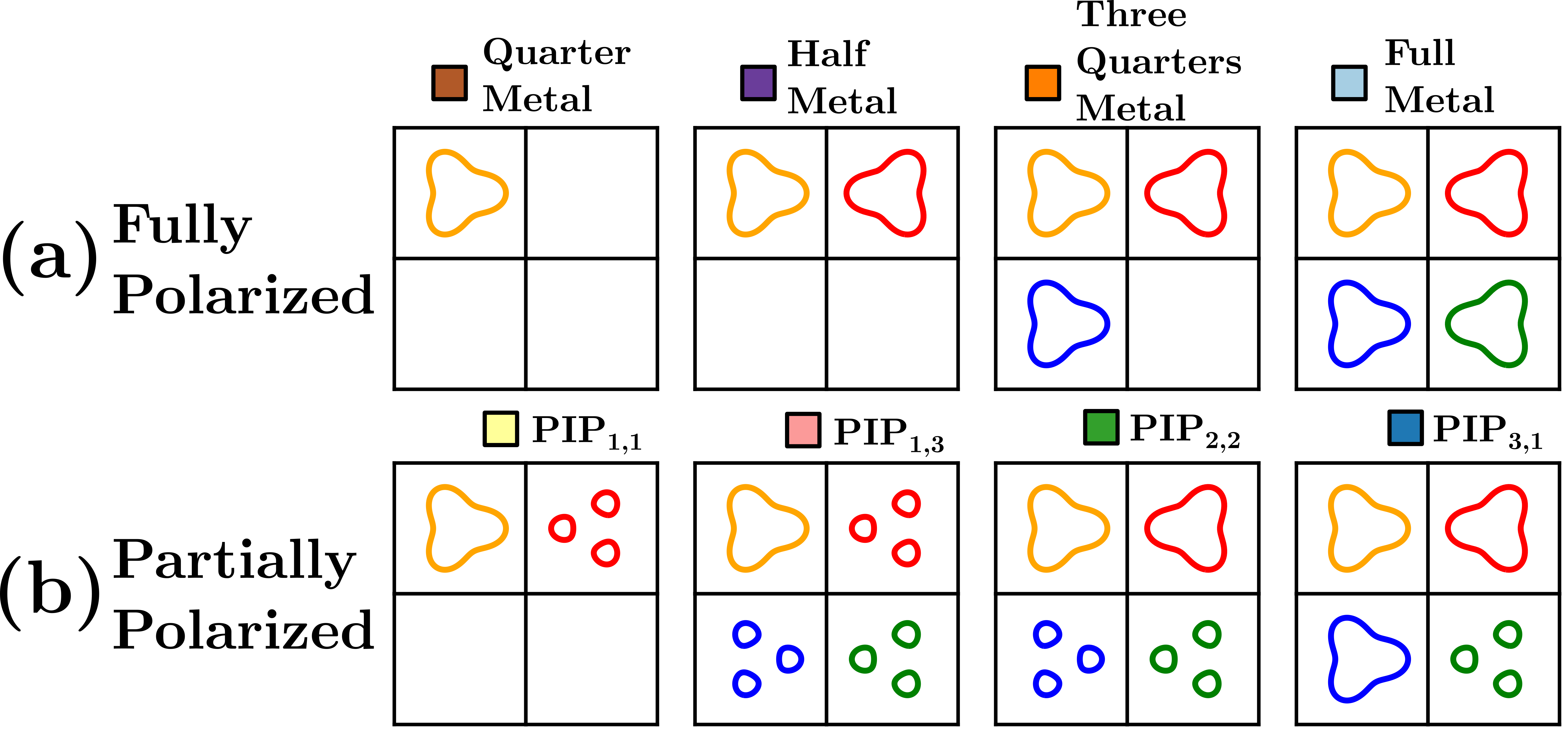}
		\caption{Schematic depictions Fermi surfaces of different phases that appear in the phase diagrams along with the color scheme used to identify them. We show (a) fully polarized and (b) partially isospin polarized states.}
		\label{fermi_surfaces}
	\end{center}
\end{figure}

\begin{figure}[h]
	\begin{center}
		\includegraphics[scale=.6]{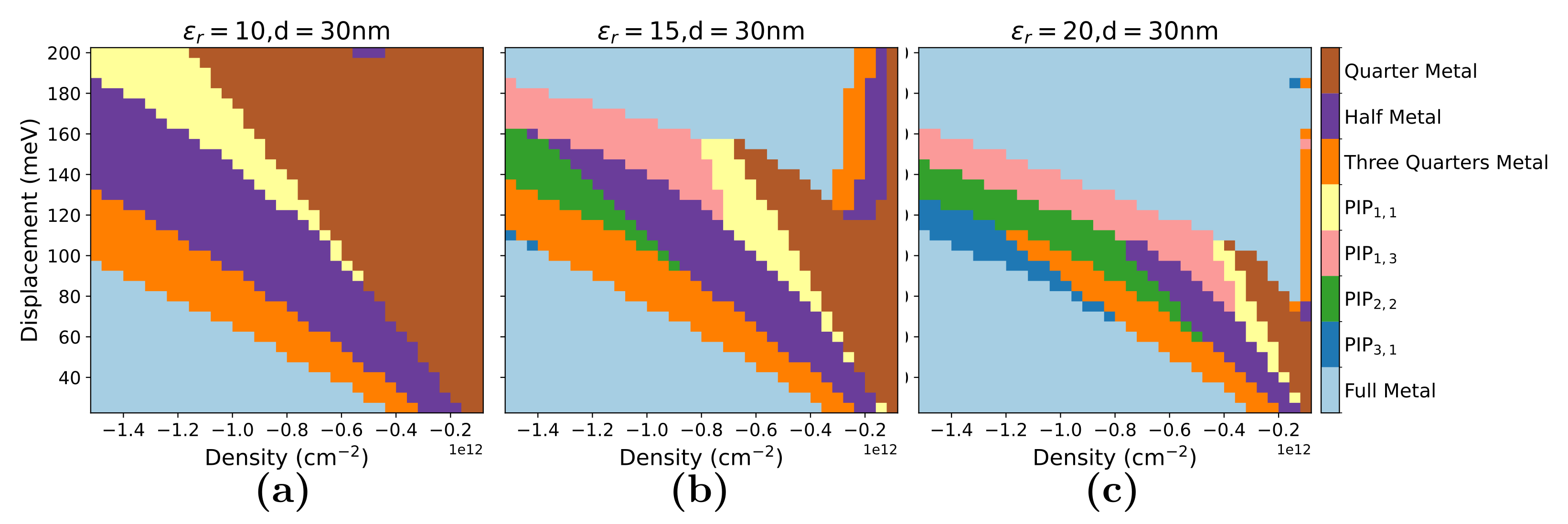}
		\caption{Phase diagram for the system with $d=30$nm. The guide to color coding is in the bar on the right.
}
		\label{d30}
	\end{center}
\end{figure}

\begin{figure}[h]
	\begin{center}
		\includegraphics[scale=.6]{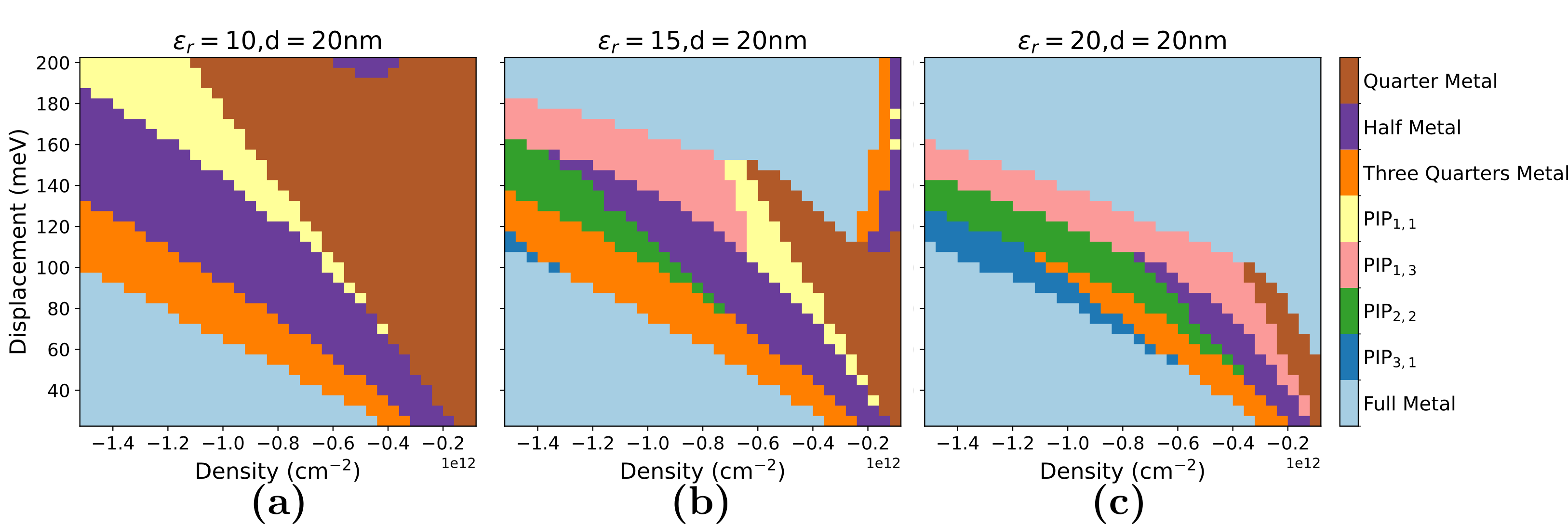}
		\caption{Phase diagram for the system with $d=20$nm. The notations are the same as in Fig. \ref{d30}.}
		\label{d20}
	\end{center}
\end{figure}

Following this procedure, we obtain the phase diagrams shown in Figs.~\ref{d30} and \ref{d20}. We generally observe a sequence of isospin-polarized states as the density of holes is increased, evolving from quarter-metal to half-metal to three-quarter-metal and finally to full metal states. Representative Fermi surfaces for these phases are shown in Fig.~\ref{fermi_surfaces}(a). This sequence is consistent with both prior theoretical work and experimental observations~\cite{zhou2022bbg,koh2024bbg,mayrhofer2025,martinez2025}. 

In addition, in between these fully polarized states, partially isospin polarized (PIP) states may appear.  The unequal occupation of isospins in the PIP phases allows for minority and majority carriers to have qualitatively different Fermi surfaces, as shown in Fig. \ref{fermi_surfaces}(b). We recall that, at high densities, the Fermi surface of one isospin in BBG in the normal state forms a single large pocket, while at low densities the system instead forms three small pockets. In the PIP states, these distinct Fermi surfaces may coexist among different isospins. Representative examples are shown in Fig.~\ref{fermi_surfaces}(b).

To explore the dependence of these phases on interaction strength and screening, we vary both the dielectric constant $\e_r$ and the gate distance $d$. Results for $d=30$ nm are shown in Fig.~\ref{d30}, where each panel corresponds to a different value of $\e_r = 10, 15,$ and $20$.
For $\e_r = 10$, shown in Fig.~\ref{d30}(a), there is a progression of quarter metal to full metal without PIP states for $D \lesssim 80$meV. However, for $D \gtrsim 80$meV, in between the quarter metal and half metal phases there is a PIP$_{1,1}$ phases, i.e. a phase with one majority carrier and one minority carrier. For $\e_r = 15$, shown in Fig.~\ref{d30}(b) there is a similar progression of states as for $\e_r = 10$. However, there are larger regions of PIP phases, as well a new PIP phases including PIP$_{1,3}$, PIP$_{2,2}$, and PIP$_{3,1}$. Lastly, for $\e_r =20$, shown in Fig. \ref{d30}(c), the PIP regions are further enlarged. There are still fully polarized phases, though they take up a smaller portion of the phase diagram than for the smaller $\e_r$. 

The results for $d=20$ nm, shown in Fig. \ref{d20}, are qualitatively similar. For $\e_r =10$, there are again quarter, half, and three-quarters metals present in the phase diagram, along with the PIP$_{1,1}$ phase. When $\e_r =15$, more PIP phases are present, and for $\e_r = 20$, even more of phases are not fully polarized. 

\begin{figure}[h]
	\begin{center}
		\includegraphics[scale=.25]{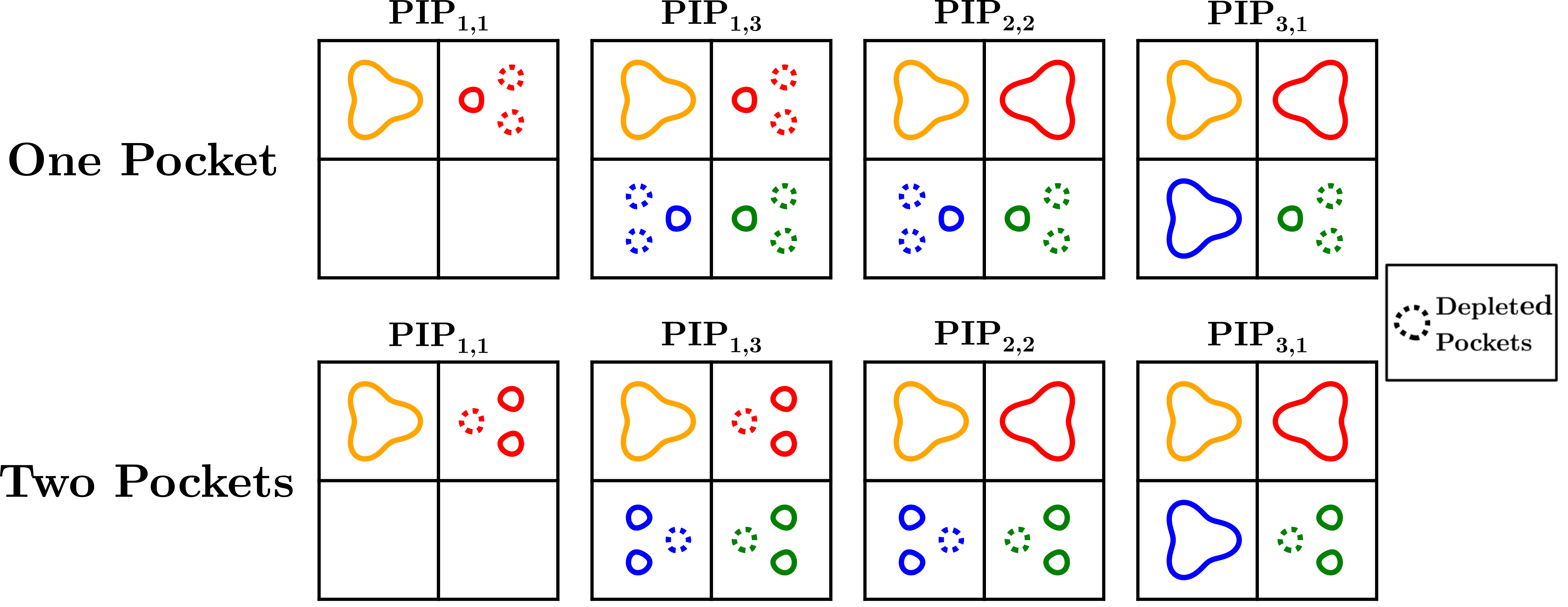}
		\caption{Schematic depictions Fermi surfaces of the different possible nematic phases. The nematic phases where the Fermi surface of the minority carriers consists of one small pocket is shown in the first row, and same is shown for the two pocket case in the second row. Dashed lines show the pockets which are expected by the $C_3$ symmetry but are depleted in the nematic phase.}
		\label{nematic_fermi_surfaces}
	\end{center}
\end{figure}

On top of these primary orders, subleading nematic orders can also develop. Particularly, in the partially isospin polarized phases, isospins may host different Fermi surfaces, and some isospins go through a nematic transition while others remain $C_3$ symmetric. In both the numerics and experimental data, the nematic states occur in the minority carriers, where density is small. Examples are shown in Fig. \ref{nematic_fermi_surfaces}, where PIP phases are depicted with minority carriers that break the $C_3$ symmetry and the majority carriers maintain this symmetry. We show several possible nematic states, however, we note that we only see nematic PIP$_{1,1}$ and PIP$_{2,2}$ in the numerical data, which is consistent with what is seen in experiment.

\begin{figure}[h]
	\begin{center}
		\includegraphics[scale=.3]{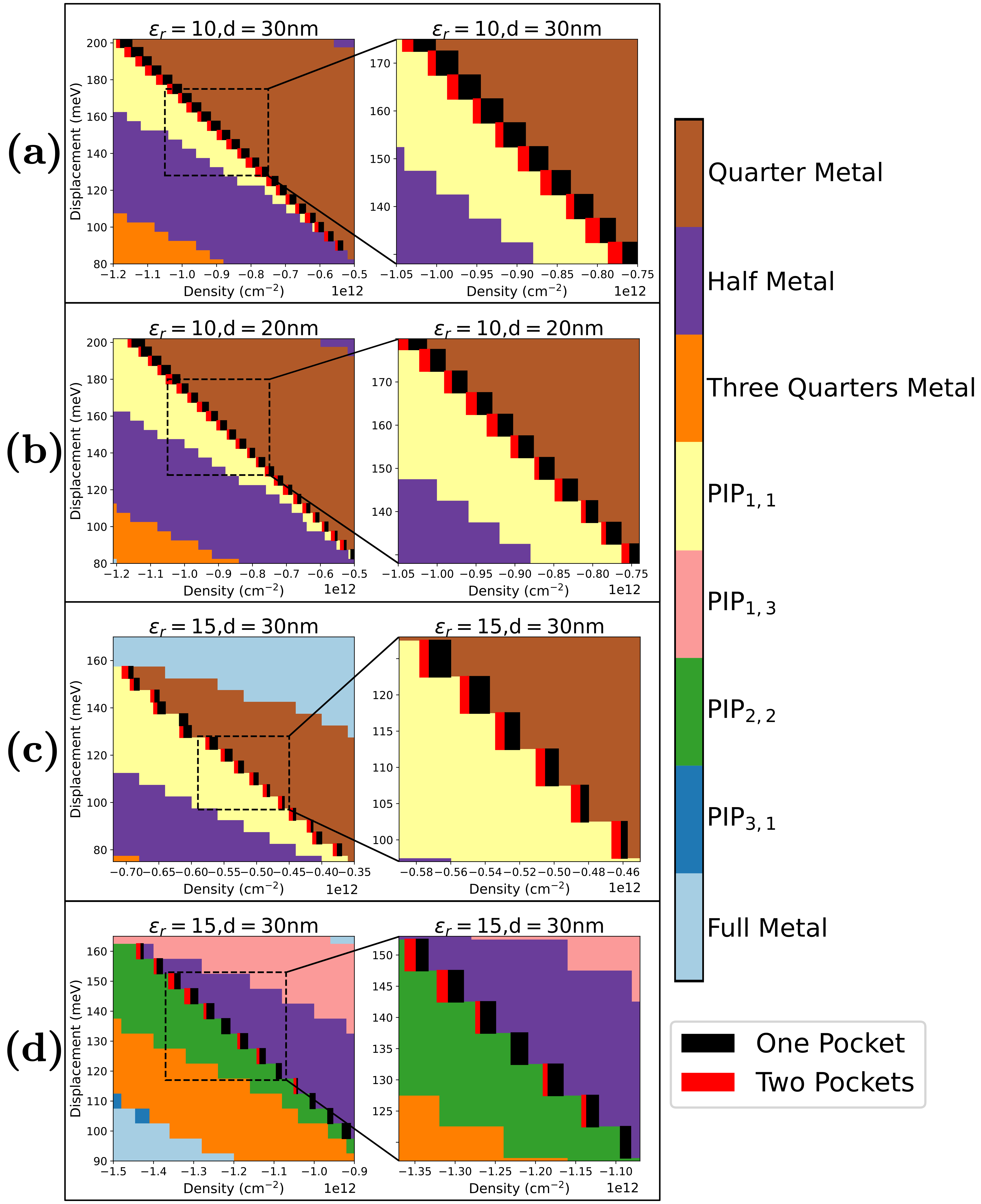}
		\caption{Regions of the phase diagrams that contain nematic phases, with the Fermi surfaces of the minority carriers consisting of one or two small pockets. In each subfigure (a)-(d), we show the full range of nematicity within the phase diagram on the left as well as a portion of the nematic states on the right for better clarity. Where the minority carriers have a single small Fermi surface are shaded with gray, and where they have two small Fermi surfaces are shaded with red.}
		\label{zoomed_phases}
	\end{center}
\end{figure}

We presented our schematic phase diagram earlier in Fig. \ref{schematic_phase_diagram}.   In Fig.  \ref{zoomed_phases} we show the actual phase diagram coming out from our numerical studies.  
We show the range of nematic states for $\e_r=10$ and $d=30$ nm in Fig. \ref{zoomed_phases}(a). Here, the nematic states form on top of the PIP$_{1,1}$ state. In this scenario, the majority carrier consists of a single large pocket, while the minority carrier has selected one or two of the original three small pockets. This nematic phase matches well with the experimentally observed phase depicted in Fig. \ref{experimental_phases}(a). The case of $\e_r=10$ and $d=20$ nm yields similar results, with nematicity appearing out of the PIP$_{1,1}$ phase.

We then show the nematic order for $\e_r = 15$ and $d=20$ nm in Fig. \ref{zoomed_phases}(c) and (d). Nematic order again appears in the PIP$_{1,1}$ state close to the quarter metal phase, as seen in Fig. \ref{zoomed_phases}(c). There is an additional region of nematic phases in PIP$_{2,2}$ state, close to the boundary of the half metal phase, shown in Fig. \ref{zoomed_phases}(d). In both cases, the structure of the nematic phase is similar: the majority carriers form a large Fermi surface, while the minority carriers occupy one or two small pockets. This nematic PIP$_{2,2}$ phase also has good agreement with the experimentally observed phases depicted in Fig. \ref{experimental_phases}(b) and (c).

We note that for $\e_r =20$ and $d = 30$ nm, no nematic states were obtained in the numerics. To confirm if this is consistent with the analytical criterion in Eq. (\ref{nematicity_condition}), both the density of states and the distance between pockets in momentum space must be computed. We do this in Appendix~\ref{effective_mass}, and find that, for $d=30$ nm, the condition on the dielectric constant for nematic order to appear is
\begin{align}
\e_r < 20.7
\end{align}
As discussed in Sec. \ref{three_pockets}, this estimate is derived in the limit of vanishing pocket size. At finite density, the pockets acquire a nonzero size. As the pocket size increases, the effective intra-pocket interaction decreases, and the condition for nematicity may no longer be satisfied at finite pocket size. Consequently, the proximity of the critical value $\e_r \approx 20$ indicates that, at $\e_r = 20$, nematic order can only occur at vanishingly small carrier densities. This explains the absence of nematic phases for this set of parameters.

We performed 
 a similar calculation for $d=20$ nm, the condition on $\e_r$ for the formation of nematic states is
\begin{align}
\e_r < 13.2.
\end{align}
For $\e_r = 15$ and $20$, the interaction is therefore too weak to stabilize nematic order, while for $\e_r = 10$ it remains sufficiently strong, leading to the observed nematic regions.

Nematic order appears only in a narrow region near the boundary between partially polarized and fully polarized states. This behavior can be understood by considering the minority carriers within the PIP phase as the density is reduced, as we showed in Sec. \ref{three_pockets}. As the minority Fermi surfaces shrink, the relevant momentum transfers involved in interactions also decrease. Since the effective intra-pocket interaction strength increases at smaller momentum transfer, the interaction increases at low densities and exceeds the critical threshold. 

\section{Conclusions}
\label{conclusions}
In this work, we have investigated emergence of nematic order in BBG. Our analysis uses the full single-particle band structure, including trigonal warping, as well as a dual-gate screened Coulomb interaction with appropriate coherence factors. By performing a self-consistent Hartree-Fock procedure, we obtained the mean-field form of the Hamiltonian. Then, by numerically solving the self-consistent equations for the order parameters, we determined the phase diagram of the system as both displacement field and number density are varied. We find that nematic phases arise near the boundaries between fully polarized and partially polarized states in the phase diagram. 
This reflects the behavior of the minority carriers, whose density is minimized near these boundaries, leading to the smallest pocket sizes. In this regime, the relevant momentum transfers are reduced, enhancing the effective interaction and enabling the onset of nematic order.
The resulting states are characterized by a reduced number of occupied pockets for the minority carriers, in qualitative agreement with experimental observations. Our results therefore provide a microscopic mechanism for the nematic behavior inferred from quantum oscillation measurements.

By varying the dielectric constant and gate distance, we further demonstrate that the stability of nematic order is strongly controlled by the interaction. Increasing $\e_r$ or reducing the gate distance weakens the effective interaction and suppresses the nematic instability. This behavior is captured quantitatively by an analytical condition derived in the low-density limit, which relates the onset of nematicity to the strength of the interaction at small momentum transfer. 

These results provide insight into the normal state from which superconductivity emerges. In Refs.~\cite{zhou2022bbg,zhang2023,holleis2025}, it was found that the normal state adjacent to superconductivity is nematic, with the minority carriers occupying only one or two pockets. Our findings therefore offer a natural microscopic framework for this state, and suggest a concrete starting point for understanding the emergence of superconductivity in BBG.

\section{Acknowledgments}
We acknowledge with thanks useful discussions with
 E. Berg,  F. Guinea, S. Nadj-Perge  and A. Young. This work was supported by U.S. Department
of Energy, Office of Science, Basic Energy Sciences, under the Award No. DE-SC0014402.
The authors acknowledge the Minnesota Supercomputing Institute at the University of Minnesota for providing computational resources.

\appendix

\section{Calculation of Effective Mass and Distance Between Pockets}
\label{effective_mass}
Here we show the calculation used to obtained the values for the effective mass of BBG at low densities. In addition, we also calculate the distance between the three pockets in momentum space. To do this, we first calculate the dispersion of larger hole band by perturbatively solving the Hamiltonian in powers of $k$. We start by writing each of the parameters in the Hamiltonian as unitless quantities. We write
\begin{align}
x_1 = \frac{D}{\gamma_1}, x_2 = \frac{\gamma_3}{\gamma_0}, x_3 = \frac{\gamma_4}{\gamma_0}, x_4 = \frac{\delta}{\gamma_1}.
\end{align}
We note that, for typical values of the tight binding parameters, that $x_1>x_2>x_3\sim x_4$. Then, the Hamiltonian written in Eq. (\ref{single_particle}) can be written as 
\begin{align}
H(\vb k)_{l l'} = \gamma_1 \begin{pmatrix}
x_1/2          & g(\vb k) & - x_3 g(\vb k)            & -x_2 g^*(\vb k) \\
g^*(\vb k) &  x_4+x_1/2           & 1    & -x_3  g(\vb k)        \\
-x_3 g^*(\vb k)            & 1       & x_4-x_1/2       &  g(\vb k)   \\
-x_2 g(\vb k)   & -x_3 g^*(\vb k)    &  g^*(\vb k) & -x_1/2             
\end{pmatrix}_{l l'},
\end{align}
where we have set $g(\vb k) = \frac{\gamma_0}{\gamma_1} f(\vb k)$.
To obtain the dispersion, we must then solve
\begin{align}
\label{det} &\text{det}\left(H(\vb k) - \e_{\vb k} I\right) = 0 = \\
\nonumber  \gamma_1^4 &\Bigg[ x_2  \left(g(\vb k)^3 + g^*(\vb k)^3 \right)\left(x_3 \left(-2 \lambda +x_3+2 x_4\right)+1\right) \\
\nonumber +&\frac{1}{4} x_1^2 \left(-2 \lambda ^2+2 \lambda  x_4+\left(x_2^2+2 x_3^2\right) |g(\vb k)|^2-x_4^2-2 |g(\vb k)|^2+1\right) 
\\
\nonumber+&\left(-\lambda ^2+\lambda +\lambda  x_4+\left(x_3-2\right) x_3 |g(\vb k)|^2+|g(\vb k)|^2\right) \left(-\lambda  (\lambda +1)+\lambda  x_4+x_3 \left(x_3+2\right) |g(\vb k)|^2+|g(\vb k)|^2\right) \\
\nonumber -&x_2^2 |g(\vb k)|^2 \left(\lambda ^2-2 \lambda  x_4+x_4^2-1\right)+\frac{x_1^4}{16} \Bigg],
\end{align}
where $\lambda = \e_{\vb k}/\gamma_1$. We first note that, close to the $K/K'$ points, $g(\vb k)$ is small and can be expanded to leading order in $\vb k$
\begin{align}
g(\vb k) \sim \frac{\sqrt{3}\gamma_0}{2\gamma_1} \tau ak e^{-i \tau \theta} = \tau z e^{-i \tau \theta},
\end{align}
where we have written $z=\frac{\sqrt{3}\gamma_0}{2\gamma_1} a k$. There is a correction of $\mathcal{O}(z^2)$ to $g$, however, this term comes with an overall prefactor of $\gamma_1/\gamma_0$, which is small. Therefore, these corrections can be safely neglected.

We then solve Eq. (\ref{det}) for $\lambda$. Since, for the upper hole band, $\e_{\vb k = 0} = -D/2$, we make an initial ansatz for the solution of $\lambda$, $\lambda = -\frac{x_1}{2} + A_1 z + A_2 z^2 + A_3 z^3 + A_4 z^4$. We then solve for each of these coefficients at each order in $z$. We also expand in powers $x_i$, and keep only leading order corrections, using $x_1>x_2>x_3\sim x_4$. We then write $\lambda$ as 
\begin{align}
\lambda = -\frac{x_1}{2} +  \left(x_1 - \frac{x_2^2}{x_1} + 2x_3 + x_4\right)z^2 - 2\frac{x_2}{x_1} \tau \cos(3\theta)z^3-\left( \frac{1}{x_1} +x_2 + 4x_3 + 3 x_4 \right)z^4
\end{align}
where we have kept $\frac{x_2^2}{x_1}$ as, for typical values of $D\sim 100$meV, this term is the same magnitude as $x_3$ and $x_4$. Eq. (\ref{disp}) is obtained by replacing the $x_i$ with the original tight-binding parameters. To calculate the effective mass of the system at low densities, we first find the location of the maximum of the band. For definitiveness, we set $\theta = \pi$ and $\tau=+1$, and solve for the maximum to find
\begin{align}
z_* = \frac{1}{\sqrt{2}} \left(x_1 + \frac{3x_2}{2\sqrt{2}} + x_3 +\frac{x_4}{2} + \frac{x_2^2}{16x_1}\right).
\end{align}
We then insert this expression for $z_*$ back into the dispersion, and expand around this point to determine the effective mass the $x$ and $y$ directions. Doing so, we obtain
\begin{align}
m_x &= \frac{8 \gamma _1 x_1}{3 \gamma _0^2 x_3^2 \left(x_1 \left(8 x_1+6 \sqrt{2} x_2+16 x_3+8 x_4\right)-x_2 \left(3 x_2+6 \sqrt{2} x_3+3 \sqrt{2} x_4\right)\right)}\\
m_y &= \frac{8 \gamma _1 x_1}{3 \gamma _0^2 x_2 x_3^2 \left(23 x_2+9 \sqrt{2} \left(2 x_1+2 x_3+x_4\right)\right)}
\end{align}
The density of states can then simply be found by taking $\nu_F = \sqrt{m_x m_y}/2\pi$. Inserting in values for the tight binding parameters, as well as a moderate value of $D = 100$meV, we find that $\nu_F \sim 8.35 \times 10^{-2}$eV$^{-1}$nm$^{-2}$.

To find the distance between the pockets $Q$, we simply note that, since the pockets are arranged symmetrically around the $K/K'$ points, we simply take the distance of the pocket from the $K/K'$ point and multiply by $\sqrt{3}$ to obtain the distance of the pockets from each other. We then obtain
\begin{align}
Q = \sqrt{3}{k_*} = \frac{2\gamma_1}{a\gamma_0}z_* = \frac{\sqrt{2} \gamma_1}{a\gamma_0} \left(x_1 + \frac{3x_2}{2\sqrt{2}} + x_3 +\frac{x_4}{2} + \frac{x_2^2}{16x_1}\right)
\end{align}
We then calculate $\tilde V = V(0) - V(Q) = \frac{e^2d}{2 \e_r \e_0}\left(1 - \frac{\tanh(dQ)}{dQ}\right)$, and find for $d = 30$nm, $\tilde V \sim 2.49\times 10^2/\e_r$eVnm$^2$, and for $d=20$nm, $\tilde V \sim 1.58\times 10^2/\e_r$eVnm$^2$. We calculate the critical $\e_r$ by taking $\nu_F \tilde V = 1$ to find, for $d=30$nm, $\e_r^c \sim 20.7$ and for $d=20$nm, $\e_r^c \sim 13.2$.
\bibliography{bbg_nematic_bib} 

\end{document}